# RAmM Algorithm(Simplex)
Subject: Cryptography

By: Jose M Manimala
under the guidance
of
Dr. Vinod Kumar P B
Head Of Department
Dept. Of Basic Sciences
RASET

# Abstract

The evolution of encryption algorithms have led to the development of very complicated and highly versatile algorithms that sacrifice efficiency for better and harder to decrypt results. But by the application of a genetic schema to the encryption of data, a new structure can be created. Genetic methods and procedures are lethal in the way they handle and manipulate data.

The RAmM algorithm uses four genetic operations that have been developed specifically for encryption of data. The operations are Replication, Augmentation, Mutation and Multiplication. The proper application of these methods according to the rules that have been found to be the best for getting optimal and correct results produces a "fingerprint" that is unique to a pair of <data , key>. This means that every single data entry can only be decrypted by using the correct set of key.

The application of the RAmM algorithm is in the field of image encryption and restoration.  The boundary and the pixel values are separately encrypted to produce a very genuine sequence that is never understood to be an image. The beauty of the procedure is that the entire image can be reproduced without any color loss or loss of pixel quality.

## INTRODUCTION

A genetic algorithm modifies data by considering data as a DNA strand and performing or adapting the same operations on the data under consideration.

A non genetic algorithm requires more complicated and intensive operations to be performed. The major advantage of a genetic algorithm is in fact the simplicity of the operation performed. Some of the more common operations seen in the DNA analysis are replication, mutation,... etc. Now consider the case where we apply these "common" observation on DNA to data by redefining the operations to perform a variety of different types of data manipulation. A cryptographic genetic algorithm is to be the topic of discussion.

## DEFINITION

The RamM(simplex) algorithm is defined as "Replicate Augment Mutate and Multiply" single direction encryption algorithm. Single direction implies only one output exists for the particular data and key. This is implied by the word simplex. A duplex version supporting multiple keys is planned as an expansion to the simplex version.

## PRELIMINARY CONSIDERATIONS

-> Both data and key have to be either of numerical or having the same pattern which is clearly quantifiable and having a measure.

-> Proportions must be defined on the data and key.

-> Data must have a layer value or a larger proportion than the key (Data>>Key).

-> Three operations are considered default and one is considered optional or special case.

    -> Replicate(R) – Default

    -> Mutate(m) – Default

    -> Multiply(M) – Default

    -> Augment(A) – Optional

## DEFINITION OF GENETIC OPERATIONS

- Replicate ---> AA
- Mutate ---> m -> A – n * B = r
- Multiply ---> A * A
- Augment ---> 01

## TERMINATING CONDITIONS

- All default operations must be applied at least once. (1)
- The algorithm terminates when (1) is satisfied and the last two operations performed must be {RXmX} replicate and mutate.

## ENCRYPTED DATA STRUCTURE

The general structure of an encrypted data packet formed using the above algorithm should be

*mX RX mX M RX ....... RXmX / remainder*

## THE ALGORITHM

1. Start
2. Check if Data > Key in proportion or value.
3. If yes then mutate the data by key
4. Check if remainder > key
5. If no them replicate and goto next step.
6. Mutate the data by key
7. Check if the remainder > key.
8. If no then Multiply. Check if remainder > key.
9. If yes Mutate the data by key.
10. If no the goto step 5.
11. After each mutate operation check the terminating condition.
12. If terminating condition is reached, generate encrypted data.
13. Stop.

## SPECIAL CASE

1. If in any case the remainder is zero and the Terminating condition has not been reached, the perform an Augment operation.
2. Then perform a replicate or mutate which ever is next in series.

## DECRYPTION

- The exact opposite of each operation is performed in reverse order.
- This reconstructs the entire data as it was.

**Example:**

DATA – 57

KEY – 5

Mode: Encryption

Check DATA > KEY – Yes

| Operation | Remainder | Encrypted Key | Terminating Condition check |
|---|---|---|---|
| Mutate | 2 | m11 | No |
| 2 < 5 | | | |
| Replicate | 22 | m11R2 | No |
| 22 > 5, Mutate | 2 | m11R2m4 | No |
| 2<5, Multiply | 4 | m11R2m4M | No |
| 4 < 5, Replicate | 44 | m11R2m4MR4 | No |
| 44 > 5, Mutate | 4 | m11R2m4MR4m8 | Yes |

Output Encrypted Key

**m11R2m4MR4m8 ~ 4**

Mode: Decryption

Key: m11r2m4mr4m8 ~ 4

| Operation | Remainder | Key | Inverse |
|---|---|---|---|
| 5 x 8 + 4 = 44 | 44 | m11R2m4Mr4 | m8 |
| 44 -> 4 | 4 | m11R2m4M | R4 |
| 4 -> 2 | 2 | m11R2m4 | M |
| 5 x 4 + 2 -> 22 | 22 | m11R2 | m4 |
| 22 -> 2 | 2 | m11 | R2 |
| 5 x 11 + 2 -> 57 | 57 | - | m11 |

**Decrypted Data: 57**

## Special Case

DATA – 55  
KEY - 5  
Mode: Encryption  
Check Data > Key - Yes

| Operation | Remainder | Encrypted Key | Terminating Condition check |
|---|---|---|---|
| Mutate | 0 | m11 | No |
| Augment | 01 | m11A | No |
| 01 < 5, Replicate | 0101 | m11AR01 | No |
| 101 > 5, Mutate | 1 | m11AR01m20 | No |
| 1 > 5, Multiply | 1 | m11AR01m20M | No |
| 1 > 5, Replicate | 11 | m11AR01m20MR1 | No |
| 11 > 5, Mutate | 1 | m11AR01m20MR1m2 | Yes |

Output Encrypted Data  
**m11AR01m20MR1m2 ~ 1**

Mode: Decryption

Key: m11AR01m20MR1m2 ~ 1

| Operation | Remainder | Key | Inverse |
|---|---|---|---|
| 5 x 2 + 1 | 11 | m11AR01m20MR1 | m2 |
| 11 -> 1 | 1 | m11AR01m20M | R1 |
| 1 -> 1 | 1 | m11AR01m20 | M |
| 5 x 20 + 1 | 101 | m11AR01 | m20 |
| 0101 | 01 | m11A | R01 |
| 01 | 0 | m11 | A |
| 5 x 11 + 0 | 55 | - | m11 |

**Decrypted Data: 55**

## Conclusion

All the special cases have been properly analyzed and examples have been given as to how the algorithm works. It can be concluded that the above algorithm is efficient and simple in its logical approach.

## References
- Cryptography Books
- Genetic Algorithms: An introduction